\documentclass[12pt, prd, showpacs]{revtex4}
\usepackage{amssymb}
\usepackage{amsmath}

\input{tcilatex}

\begin{document}

\title{Zero-momentum trajectories inside a black hole and high energy
particle collisions}
\author{A. V. Toporensky}
\affiliation{Sternberg Astronomical Institute, Lomonosov Moscow State University,
Universitetsky Prospect, 13, Moscow 119991, Russia}
\email{atopor@rambler.ru}
\affiliation{Kazan Federal University, 18 Kremlyovskaya St., Kazan 420008, Russia}
\author{O. B. Zaslavskii}
\affiliation{Department of Physics and Technology, Kharkov V.N. Karazin National
University, 4 Svoboda Square, Kharkov 61022, Ukraine}
\email{zaslav@ukr.net}
\affiliation{Institute of Mathematics and Mechanics, Kazan Federal University, 18
Kremlyovskaya St., Kazan 420008, Russia }

\begin{abstract}
We consider properties of the trajectory with the zero momentum inside a
spherically symmetric black hole. We work mostly in the Painlev\'{e}%
-Gullstrand frame and use the concept of the "river model of black hole".
This consept allows us to decompose (in a "cosmological manner") the
geodesic motion of a test particle into a "flow" of the frame and a peculiar
motion with respect to this frame. After this decomposition the application
of standard formulae of special relativity for kinematic processes becomes
possible. The present paper expands the notion of peculiar velocities to the
region under the horzion and exploits it for the description of two physical
processess - high energy collisions and redshift. Using this approach we (i)
present a novel description of particle collisions occuring near black hole
horizons inside the event horizon. In particular, we show that the
trajectory under discussion is relevant for ultra-high energy collisions.
(ii) In the framework of the river model, we derive a simple formula (both
outside and inside the horizon) for the redshift in the case of radial
motion. It represents the product of two factors. One of them is responsible
for pure gravitational part whereas the other one gives the Doppler shift
due to peculiar motion with respect to the "flow".
\end{abstract}

\maketitle

\section{Introduction}

The motivation of the present paper is two-fold. In recent years, a new
approach to the description of kinematics of particles moving in the
background of a black hole was suggested. This is so-called the "river model
of black hole" \cite{ham}. It gives a quite clear presentation of particle
motion outside the horizon. Formally, it is quite correct also inside but it
deals there with the quantities that have no direct meaning (for example,
the coordinate velocity of the flow that becomes superluminal there). In the
present article, we elaborate the approach that is based on the combination
of this model with the notion of peculiar \ velocity. This enables us to
interpret all processes inside a black hole in terms of physical quantities.
As the metric inside the Schwarzschild-like black hole is essentially
nonstationary, the general approach under discussion can be also useful in
cosmological problems.

Although, in principle, any physical result can be obtained with more
standard means, without invoking the river model, this approach is useful at
least for two reasons. First, it enables one to carry out analogy with the
cosmological problems where the notions of the Hubble flow and peculiar
velocities (that measure deviation from it) are rather simple and clear.
Second, it turns out that velocities under discussion have direct meaning,
being tetrad components measured by a falling observer. As a result, one can
exploit the formulas typical of the locally flat background thus using the
results of special relativity.

We also apply this formalism to the description of high energy particle
collisions inside black holes that is another motivation of our work. In
last decade, much attention was paid to such collisions near black holes.
This direction is stimulated by observation made in \cite{ban} according to
which the energy $E_{c.m.}$ in the centre of mass frame of colliding
particles can grow unbounded under some additional conditions. These
conditions imply that for one of two particles the relationship between the
energy and angular momentum or electric charge is fine-tuned (the
corresponding trajectory is called critical). This is the so-called Ba\~{n}%
ados-Silk-West (BSW) effect. After these findings, the interest to earlier
works also revived \cite{pir1}, \cite{pir3}. The emphasis in subsequent
researches was made on collisions outside the horizon close to it.
Meanwhile, there is another, quite subtle (sometimes even contradictory)
issue concerning high energy collisions inside a black hole \cite{lake} - 
\cite{inner}. In doing so, such collisions were mainly considered for the
Reissner-Nordstr\"{o}m (RN) or Kerr black hole near the inner horizon.

In the present paper, we draw attention to the fact that there exists
another version of high energy collisions that is possible even inside the
Schwarzschild metric. The role of the critical trajectory is played by the
particle with a zero radial momentum. It extends the class of trajectories
in a strong gravitational field for which some of the components of the
momentum are equal to zero - cf. the zero-angular momentum observers (ZAMO) 
\cite{72} or zero energy observers \cite{01}, \cite{02}. However, in
contrast to the aforementioned cases, there is no analogue of the trajectory
under discussion outside the horizon.

One reservation is in order. It is known that the full structure of the
inner region of a charged dynamically formed black hole can essentially
differ from the textbook description of the RN black hole (see, e.g. \cite%
{brady}). However, this problem does not arise for noncharged black holes
that are the main subject of our discussion. But even for the RN black hole,
our consideration retains at least methodical value since it gives a
detailed description of properties of peculiar velocities in different
space-times that can be useful in more realistic contexts as well. We would
like to stress also, that as a black hole is an object, important both for
theoretical physics and astronomy, investigation of all its structure
including the inner region is physically relevant.

Throughout the paper, we use geometric units in which fundamental constants $%
G=c=1$.

\section{Basic equations}

\subsection{Outside a black hole}

We consider the metric%
\begin{equation}
ds^{2}=-fdt^{2}+\frac{dr^{2}}{f}+r^{2}d\omega ^{2}\text{,}  \label{met}
\end{equation}%
where $d\omega ^{2}=(d\theta ^{2}+\sin ^{2}\theta d\phi ^{2}).$

We suppose that the metric has the event horizon at $r=r_{+}$, so $%
f(r_{+})=0 $. For the Schwarzschild metric, $f=1-\frac{r_{+}}{r}$, where $%
r_{+}=2M$ is the horizon radius, $M$ being the black hole mass. The most
part of results applies also to generic $f(r)$. We will also discuss briefly
metrics with the inner horizon $r_{-}<r_{+}$, $f(r_{-})=0$. Near the event
horizon,

\begin{equation}
f\approx 2\kappa (r-r_{+})\text{,}  \label{kar}
\end{equation}%
where $\kappa =\frac{f^{\prime }(r_{+})}{2}$ is the surface gravity. In the
Schwarzschild case $\kappa =\frac{1}{2r_{+}}$.

Let us consider geodesic motion of a massive particle (we call it
"observer"). The geodesic equations of motion within the equatorial plane
for such a particle read:%
\begin{equation}
\dot{t}=-\frac{\varepsilon }{f}\text{,}  \label{mt}
\end{equation}%
\begin{equation}
\dot{\phi}=\frac{L}{mr^{2}}\text{,}
\end{equation}%
\begin{equation}
\dot{r}=\sigma Z\text{, }Z=\sqrt{\varepsilon ^{2}-f(1+\frac{L^{2}}{r^{2}m^{2}%
})}\text{,}  \label{mr}
\end{equation}%
$E$ is the conserved energy $E=-mu_{t}$ of a particle, the four-velocity $%
u^{\mu }=\frac{dx^{\mu }}{d\tau }$, dot denotes derivative with respect to
the proper time $\tau ,$ $L$ being the angular momentum, $\varepsilon =\frac{%
E}{m},$ $\sigma =\pm 1$ depending on the direction of motion.

In eqs. (\ref{mt}), (\ref{mr}) it is implied that $f>0$, so motion occurs in
the outer region of a black hole. For the four-velocity $u^{\mu }$ we have
from (\ref{mt}) - (\ref{mr}) in coordinates $(t,r,\phi )$:%
\begin{equation}
u^{\mu }=(\frac{\varepsilon }{f}\text{, }\sigma Z\text{, }\frac{L}{mr^{2}})%
\text{.}
\end{equation}

\subsection{Inside a black hole}

Inside a black hole, we can choose $r=-T$, $t=y$, $f=-g$, $g>0$. Then,%
\begin{equation}
ds^{2}=-\frac{dT^{2}}{g}+gdy^{2}+T^{2}d\omega ^{2}\text{.}  \label{min}
\end{equation}

For the Schwarzschild metric $g=\frac{r_{+}}{r}-1=-1-\frac{r_{+}}{T},$ where 
$-r_{+}<T\leq 0$ (see \cite{nov1} and \cite{fn}, page 25).

As the metric does not depend on $y$, the radial momentum $\varepsilon
=u_{y} $ is conserved. Two equations of motion for a geodesic particle
within the plane $\theta =\frac{\pi }{2}$ read%
\begin{equation}
\dot{y}=-\frac{\varepsilon }{g}\text{,}  \label{my}
\end{equation}%
\begin{equation}
m\dot{\phi}=\frac{L}{T^{2}}\text{.}
\end{equation}%
Here, $\varepsilon $ can have any sign, $\varepsilon =\pm \left\vert
\varepsilon \right\vert .$ The case $\varepsilon =0$ is also possible. From
the normalization \ condition $u_{\mu }u^{\mu }=-1$ and taking into account
the forward-in-time condition $\dot{T}>0$, we obtain

\begin{equation}
\dot{T}=Z\text{, }  \label{mT}
\end{equation}%
\begin{equation}
Z=\sqrt{\varepsilon ^{2}+g(1+\frac{L^{2}}{r^{2}m^{2}})}\text{.}  \label{zt}
\end{equation}%
Thus in coordinates $(T,y,\phi )$ we have \ for the four-velocity of a
particle:%
\begin{equation}
u^{\mu }=(Z,\frac{\varepsilon }{g},\frac{L}{mT^{2}})\text{,}  \label{muu}
\end{equation}%
\begin{equation}
u_{\mu }=(-\frac{Z}{g},\varepsilon ,\frac{L}{m}).  \label{mul}
\end{equation}

\section{Massless case}

The above formulas apply to massive particles. Below, we list the similar
ones for motion of massless particles (for example, photons). In a similar
way, the components $k_{t}=-\omega _{0}$, $k_{\phi }=l$ of the wave vector
are conserved. The normalization condition $k_{\mu }k^{\mu }=0$ gives us
outside the event horizon%
\begin{equation}
k^{\mu }=(\frac{\omega _{0}}{f},\sigma z\text{,}\frac{l}{r^{2}})\text{, }%
k_{\mu }=(-\omega _{0}\text{, }\frac{\sigma z}{f}\text{, }l)\text{,}
\label{kf}
\end{equation}%
where,%
\begin{equation}
z=\sqrt{\omega _{0}^{2}-\frac{fl^{2}}{r^{2}}}\text{.}  \label{kl}
\end{equation}

Under the horizon, we have in the coordinates $(T$, $y$, $\phi )$ 
\begin{equation}
k^{\mu }=(z,\frac{\sigma \omega _{0}}{g},\frac{l}{T^{2}})\text{,}  \label{k}
\end{equation}%
\begin{equation}
k_{\mu }=(-\frac{z}{g},\sigma \omega _{0},l)\text{,}
\end{equation}%
where now $k_{y}=\pm \omega _{0}$ is conserved, $\omega _{0}>0$,%
\begin{equation}
z=\sqrt{\omega _{0}^{2}+\frac{g}{T^{2}}l^{2}}\text{.}  \label{z}
\end{equation}

In what follows, we are mainly interested in radial motion only, so we put $%
l=0=L$. Then,%
\begin{equation}
Z=\sqrt{\varepsilon ^{2}-f}=\sqrt{\varepsilon ^{2}+g}\text{, }z=\omega _{0}.
\label{Zz}
\end{equation}

\section{Gullstrand-Painlev\'{e} frame and tetrads}

In what follows, we will use the so-called generalized Gullstrand-Painlev%
\'{e} form (GP) of the metric and the concept of "river of space" related to
it \cite{ham}. For a generic $f\,$, the transformation where%
\begin{equation}
\tilde{t}=t+\int^{r}\frac{dr^{\prime }}{f}v\text{,}  \label{tt}
\end{equation}

\begin{equation}
v=\sqrt{1-f}\text{,}  \label{vf}
\end{equation}%
brings the metric to the form, nonsingular on the horizon \cite{ham}, \cite%
{3}. For the Schwarzschild case, this reduces to the Gullstrand-Painlev\'{e}
metric \cite{pain}, \cite{gull}.

The metric can be written as%
\begin{equation}
ds^{2}=-d\tilde{t}^{2}+(dr+vd\tilde{t})^{2}+r^{2}d\omega ^{2}\text{.}
\label{gp}
\end{equation}

It is regular in the vicinity of the horizon, where $f=0$. The velocity $v$
has a simple physical meaning \cite{ham}. This is a velocity of "flow"
because of a "river of space". In the frame (\ref{gp}) the proper distance
on the cross-section $\tilde{t}=const$ between different points 1 and 2 is
equal to $\left\vert r_{2}-r_{1}\right\vert $. Therefore, $v\,\ $measures
the rate with which such a proper distance changes with time. This
interpretation is applicable even under the horizon in spite of the fact
that the vector normal to the hypersurface $r=const$ is time-like there, so $%
r$ itself is a time-like coordinate.

For a \ particle with arbitrary $\varepsilon =\frac{E}{m}$ moving with
decreasing of $r$, equations of motion (\ref{mt}), (\ref{mr}) give us%
\begin{equation}
\frac{dr}{d\tau }=-Z\text{, }Z=\sqrt{\varepsilon ^{2}+v^{2}-1}\text{,}
\label{zv}
\end{equation}

where we used (\ref{vf}). From (\ref{mt}) and (\ref{tt}) we have%
\begin{equation}
\frac{d\tilde{t}}{d\tau }=\frac{\varepsilon }{f}-\frac{v}{f}Z.  \label{ttil}
\end{equation}

Taking into account (\ref{tt}) we have%
\begin{equation}
\frac{dr}{d\tilde{t}}=\frac{Zf}{(Zv-\varepsilon )}.  \label{rtil}
\end{equation}

\section{Choice of tetrads}

In what follows, it is convenient to introduce tetrads since this enables us
to analyze local events in a way similar to special relativity. We choose
tetrads in such a way that $h_{(0)}^{\mu }$ coincides with the four-velocity
of an observer comoving with the flow. Such an observer has $\varepsilon =1$%
. Then, it follows from (\ref{vf}), (\ref{zv}), (\ref{rtil}) that $\frac{dr}{%
d\tilde{t}}=-v$ for him. Explicitly,

\begin{equation}
h_{(0)}^{\mu }=\frac{\partial }{\partial \tilde{t}}-v\frac{\partial }{%
\partial r}\text{.}
\end{equation}%
For the rest of tetrads we choose

\begin{equation}
h_{(1)}^{\mu }=\frac{\partial }{\partial r}\text{, }h_{(2)}^{\mu }=\frac{1}{r%
}\frac{\partial }{\partial \theta }\text{, }h_{(3)}^{\mu }=\frac{1}{r\sin
\theta }\frac{\partial }{\partial \phi }.
\end{equation}

For a generic particle, it is also convenient to split this quantity to $-v$
(the flow velocity) and the peculiar velocity $v_{p}$, so by definition%
\begin{equation}
\frac{dr}{d\tilde{t}}=-v+v_{p}\text{.}  \label{sp}
\end{equation}

We can also define the three-velocity in a standard way(see, e.g. eq. 3.9 of 
\cite{72}):%
\begin{equation}
V^{(i)}=-\frac{h_{(i)\mu }u^{\mu }}{h_{(0)\mu }u^{\mu }}.
\end{equation}%
It is easy to find that for our choice of tetrads,%
\begin{equation}
-h_{(0)\mu }u^{\mu }=u^{\tilde{t}}
\end{equation}%
and%
\begin{equation*}
h_{(1)\mu }u^{\mu }=vu^{\tilde{t}}+u^{r}\text{.}
\end{equation*}%
Then, we obtain for pure radial motion 
\begin{equation}
V^{(1)}=v+\frac{dr}{d\tilde{t}}=v_{p}.  \label{V}
\end{equation}%
Thus the tetrad component of the three-velocity coincides with the peculiar
one that agrees with \cite{ham}.

\section{Properties of the peculiar velocity}

Now, we will discuss separately the properties of $v_{p}$ in the outer
region and inside the horizon.

\subsection{Outside the horizon}

It is known \cite{k} that in the outer region

\begin{equation}
\varepsilon =\frac{\sqrt{f}}{\sqrt{1-V^{2}}}=\frac{\sqrt{1-v^{2}}}{\sqrt{%
1-V^{2}}},  \label{ef}
\end{equation}

where $V$ is the velocity measured by a static observer with tetrads

\begin{equation}
h_{(0)st}^{\mu }=\frac{1}{\sqrt{f}}\frac{\partial }{\partial t}\text{, }%
h_{(1)st}^{\mu }=\sqrt{f}\frac{\partial }{\partial r}\text{.}
\end{equation}

It follows from (\ref{vf}), (\ref{rtil}), (\ref{sp}), (\ref{zv}) that

\begin{equation}
v_{p}=\frac{v-V}{1-vV},  \label{vV}
\end{equation}%
\begin{equation}
V=\frac{v-v_{p}}{1-vv_{p}}\text{,}  \label{V(v)}
\end{equation}%
where%
\begin{equation}
V=\frac{Z}{\varepsilon }\text{, }0\leq V\leq 1\text{.}  \label{st}
\end{equation}%
Actually, this is a relativistic formula for transformation of velocities.
It is clear from (\ref{mr}), (\ref{zv}) that $V\leq 1$. As $v\leq 1$ also,
we see that $\left\vert v_{p}\right\vert \leq 1$ as it should be for the
concept of the river of space \cite{ham} to be self-consistent.

It is seen from (\ref{vV}) that%
\begin{equation}
v_{p}-v=-\frac{V(1-v^{2})}{1-vV}\leq 0\text{.}  \label{v-v}
\end{equation}

If $\varepsilon >1$, it is easy to check that $V>v$, so it follows from (\ref%
{vV}) that $v_{p}<0$. If $\varepsilon <1$, $v>V$ and $v_{p}>0$. If $%
\varepsilon =1$, $v=V$ and $v_{p}=0$.

Thus%
\begin{equation}
signv_{p}=sign(1-\varepsilon )\text{,}  \label{1-e}
\end{equation}

It follows from, (\ref{ef}), (\ref{V(v)}), (\ref{st}) that%
\begin{equation}
\varepsilon =\frac{1-vv_{p}}{\sqrt{1-v_{p}^{2}}}.  \label{eps}
\end{equation}

The inverse formula reads%
\begin{equation}
v_{p}=\frac{v}{v^{2}+\varepsilon ^{2}}-\frac{\varepsilon Z}{%
v^{2}+\varepsilon ^{2}}\text{,}  \label{inv}
\end{equation}%
where $Z$ is given by (\ref{zv}). (Formally, there is also the root of eq. (%
\ref{eps}) with the plus sign but it does not satisfy the condition $v_{p}=0$
if $\varepsilon =1$. One can show that eq. (6.9) is valid also for negative $%
\varepsilon $.)

Using (\ref{mt}), (\ref{sp}), we obtain%
\begin{equation}
\varepsilon =f\frac{d\tilde{t}}{d\tau }-v\frac{dr}{d\tau }=\frac{d\tilde{t}}{%
d\tau }(1-vv_{p})\text{,}  \label{epsf}
\end{equation}%
whence%
\begin{equation}
\frac{d\tilde{t}}{d\tau }=\frac{1}{\sqrt{1-v_{p}^{2}}}\text{.}  \label{dt}
\end{equation}

Then, after straightforward algebraic manipulations we find for a fixed $%
\varepsilon $ from (\ref{eps})%
\begin{equation}
v_{p}^{2}\frac{dv}{dv_{p}}=\frac{v_{p}(v_{p}-v)}{1-v_{p}^{2}}\text{.}
\label{dv}
\end{equation}

Taking into account (\ref{v-v}) and (\ref{1-e}), we see that $\frac{dv_{p}}{%
dv}>0$ for $\varepsilon >1$, $\frac{dv_{p}}{dv}=0$ for $\varepsilon =1$ and $%
\frac{dv_{p}}{dv}<0$ for $\varepsilon <1$.

It is instructive to rewrite eq. (\ref{dv}) in the form%
\begin{equation}
\frac{dv_{p}^{2}}{dv}=\frac{2v_{p}^{2}(1-v_{p}^{2})}{v_{p}-v}.  \label{vp2}
\end{equation}%
Taking into account (\ref{v-v}) we see that%
\begin{equation}
\frac{dv_{p}^{2}}{dv}\leq 0\text{,}  \label{dvv}
\end{equation}%
whence%
\begin{equation}
sign\frac{dv_{p}^{2}}{dr}=sign\frac{df}{dr}  \label{vpf}
\end{equation}%
independently of $\varepsilon $.

\subsection{Inside the event horizon}

Here, eq. (\ref{ef}) is somewhat modified to give%
\begin{equation}
\varepsilon =\sigma \frac{\sqrt{g}V}{\sqrt{1-V^{2}}}=\sigma \frac{\sqrt{%
v^{2}-1}V}{\sqrt{1-V^{2}}}\text{,}  \label{ein}
\end{equation}%
see \cite{inner}, eq. (28). The factor $V$ typical of the momentum appears
due to the fact that under the horizon the integral of motion has the
meaning of momentum (not energy), now instead of (\ref{st})%
\begin{equation}
V=\frac{\left\vert \varepsilon \right\vert }{Z}\leq 1\text{.}  \label{vez}
\end{equation}

The velocity $V$ is measured with respect to an observer who remains at rest
under the horizon in the sense that $y=const$, $\varepsilon =0$. Such a
geodesic trajectory does not have analogues outside the horizon \cite{lobo}.

Then, repeating formulas step by step, we obtain%
\begin{equation}
v_{p}=\frac{\tilde{v}-\sigma V}{1-\tilde{v}\sigma V},  \label{vpv}
\end{equation}%
where 
\begin{equation}
\tilde{v}=\frac{1}{v}=\frac{1}{\sqrt{1+g}}\leq 1\text{.}  \label{vtil}
\end{equation}%
If $\varepsilon =0$, $Z=\sqrt{-f}$, and we have from (\ref{rtil}) that%
\begin{equation}
\frac{dr}{d\tilde{t}}=\frac{f}{v}\text{,}
\end{equation}%
\begin{equation}
v_{p}=\tilde{v}=\frac{1}{v}=\frac{1}{\sqrt{1+g}}\text{,}  \label{vv}
\end{equation}%
where we used (\ref{vf}). Eq. (\ref{vv}) is valid for any point inside the
horizon. Now,%
\begin{equation}
V=0\text{.}  \label{v0}
\end{equation}

One can see that%
\begin{equation}
sign(\tilde{v}-v_{p})=\sigma =sign\varepsilon ,  \label{sign}
\end{equation}%
eq. (\ref{dv}) is still valid. Using eq. (\ref{vf}), it is possible to
rewrite eq. (\ref{dv}) in the form of differential equation for $dv_{p}/dr$
for any particular $f(r)$ and use it, if needed, to find $v_{p}(r)$ if the
initial value of peculiar velocity is given. It is easy to check that eq. (%
\ref{1-e}) is valid as well.

Bearing also in mind (\ref{v-v}), we can write the general result as%
\begin{equation}
sign\frac{dv_{p}}{dv}=sign(\varepsilon -1)\text{.}  \label{sgn}
\end{equation}

Eqs. (\ref{sgn}), (\ref{1-e}), (\ref{vp2}), (\ref{dvv}) are valid both
outside and inside the horizon.

If $\varepsilon =1$, it follows from (\ref{zv}), (\ref{vez}) that outside
the horizon $v_{p}=0$. Inside the horizon, the same conclusion follows from (%
\ref{vpv}) where now $\sigma =+1$ since $\varepsilon $ in (\ref{ein}) is
positive. Thus $v_{p}=0,$ $\frac{dv_{p}}{dv}=0$ both for the regions outside
and inside the horizon.

It is also instructive to note a useful formula%
\begin{equation}
Z=\frac{v-v_{p}}{\sqrt{1-v_{p}^{2}}}  \label{zvp}
\end{equation}%
that is valid both outside the horizon and inside. It can be obtained from (%
\ref{Zz}) and (\ref{eps}). It tells us that the radial momentum arises due
to motion of a particle against the flow, with the Lorentz factor taken into
account. It is seen from (\ref{v-v}) that Z is nonnegative as it should be.

\section{The horizon limit}

We are interested in what happens near the horizon. Let us consider
separately different cases depending on the sign of $\varepsilon $.

\subsection{$\protect\varepsilon >0$}

Now, it is easy to check that near the horizon,%
\begin{equation}
Z\approx \varepsilon -\frac{f}{2\varepsilon }\text{.}  \label{zf}
\end{equation}%
It follows from (\ref{rtil}), (\ref{sp}) that%
\begin{equation}
\frac{dr}{d\tilde{t}}\approx -\frac{2\varepsilon ^{2}}{\varepsilon ^{2}+1}%
\text{,}
\end{equation}

\begin{equation}
v_{p}\rightarrow \frac{1-\varepsilon ^{2}}{1+\varepsilon ^{2}}.  \label{vpe}
\end{equation}%
As far as $V$ is concerned, we have from (\ref{st}), (\ref{zf}) that \ in
the horizon limit

\begin{equation}
V\approx 1-\frac{f}{2\varepsilon ^{2}}.  \label{et}
\end{equation}

\subsection{$\protect\varepsilon =0$}

In the horizon limit $g\rightarrow 0$ we obtain from (\ref{vv}), that%
\begin{equation}
v\rightarrow 1,\text{ }v_{p}\rightarrow 1\text{.}  \label{vhor}
\end{equation}%
According to (\ref{v0}), $V=0$.

\subsection{$\protect\varepsilon <0$}

This case can be realized under the event horizon only, where $f<0$.

Using (\ref{Zz}) and (\ref{rtil}), we obtain

\begin{equation}
Z\approx -\varepsilon +O(f)\text{,}  \label{zf2}
\end{equation}%
\begin{equation}
\frac{dr}{d\tilde{t}}\approx \frac{f}{2}\rightarrow -0,
\end{equation}%
Then, we have from (\ref{sp}) that%
\begin{equation}
v_{p}=1-\frac{f^{2}}{8\varepsilon ^{2}}+O(f^{3})\rightarrow 1\text{.}
\label{pneg}
\end{equation}

Under the horizon, it follows from (\ref{ein}) that again $V\rightarrow 1$.

\section{Dependence of $v_{p}$ on time}

It is convenient to consider, how the peculiar velocity changes during
particle motion, for different cases separately.

\subsection{$\protect\varepsilon >1$}

Now $v_{p}<0$, $\frac{dv_{p}}{dv}>0$ according to eqs. (\ref{1-e}), (\ref%
{sgn}). Let a particle moving from the outer region cross the event horizon.
Outside the horizon, the quantity $v$ (\ref{vf}) increases monotonically
from $v=0$ at infinity to $v=1\,\ $\ at the horizon. In doing so, the
peculiar velocity $v_{p}$ increases from $-\sqrt{1-\frac{1}{\varepsilon ^{2}}%
}$ at infinity to the value $v_{p}(r_{+})<0$ (\ref{vpe}) on the horizon.

After crossing the horizon, the picture depends on the type of metric. For
the Schwazrschild metric, after crossing the event horizon, the metric
coefficient $g$, the quantity $v=\sqrt{1+g}$ and $v_{p}$ continue to
increase monotonically. When the $r=0$ singularity is approached, $%
v\rightarrow \infty $, $V\rightarrow 0$ and $v_{p}\rightarrow -0$, as it
follows from (\ref{vez}), (\ref{vpv}).

For the RN metric, the function $g$ attains its maximum value in the point 
\begin{equation}
r_{0}=\frac{2r_{+}r_{-}}{r_{+}+r_{-}},  \label{r0}
\end{equation}%
\begin{equation}
g(r_{0})=\frac{(r_{+}-r_{-})^{2}}{4r_{+}r_{-}}.  \label{g0}
\end{equation}%
Therefore, $v_{p}$ increases in the interval $(r_{+},r_{0})$, where it
attains the maximum value in the same point $r_{0}$. Further, $g$ and $v$
begin to decrease and so does $v_{p}$. On the inner horizon $r=r_{-}$, $v=1$
and (\ref{vpe}) with $v_{p}(r_{-})=v_{p}(r_{+})<0$ is valid.

\subsection{$\protect\varepsilon =1$,}

According to what is said in the paragraph after eq. (\ref{v-v}), $%
v_{p}=const=0$ now. $\ $

\subsection{$0<\protect\varepsilon <1$\ }

Now $v_{p}>0$, $\frac{dv_{p}}{dv}<0$, according to eqs. (\ref{1-e}), (\ref%
{sgn}). For the Schwarzschild metric, $v_{p}$ under the event horizon
monotonically decreases. It is seen from (\ref{vpe}) that on the horizon, $%
v_{p}(r_{+})$\thinspace $>0$. Near the singularity, $v\rightarrow \infty ,$ $%
V\rightarrow 0$ and $v_{p}\rightarrow +0$. For the RN metric, $v_{p}$
decreases, attains its minimum value at $r_{0}$, afterwards it increases. On
the inner horizon, $v=1$, $v_{p}(r_{-})=v_{P}(r_{+})>0$ according to (\ref%
{vpe}).

It is instructive to extract from the above results the behavior of the
absolute value $\left\vert v_{p}\right\vert $ in comparison with $V$. In the
outer region, $\left\vert v_{p}\right\vert $ decreases ($V$ increases to $1$
on the event horizon). Inside the event horizon, the behavior is different
for different types of metric. For the Schwarzschild one, $\left\vert
v_{p}\right\vert $ continues to decrease, $\left\vert v_{p}\right\vert =0$
in the singularity $r=0$ ($V$ also decreases to $0$ in the singularity)
independently of $\varepsilon $. In the RN metric, $\left\vert
v_{p}\right\vert $ continues to decrease from $r_{+}$ to $r_{0}$, afterwards
it increases ($V$ so does, $V=1$ on each horizon). In the region between the
singularity and $r_{-}$, the notion of the velocity of flow loses its sense
since there is a point $r_{1}$ there such that $f(r_{1})=1.$ Then, for $%
0\leq r<r_{1}$the velocity of flow (\ref{vf}) becomes formally imaginary.

Finally, we should note that the case of $\varepsilon \leq 0$ is available
only inside the event horizon. The behavior of $v_{p}$ is the same as for $%
0<\varepsilon <1$ except for the asymptotics near the inner horizon where
now $v_{p}\rightarrow 1$.

All the features described above can be also obtained directly from eq. (\ref%
{inv}). It is instructive to summarize in tables the corresponding results
for the properties of peculiar veocities described above. See Table 1 and
Table 2 below.

Table 1. General features

\begin{tabular}{|l|l|l|l|l|}
\hline
& $\varepsilon >1$ & $\varepsilon =1$ & $0<\varepsilon <1$ & $\varepsilon
\leq 0$ \\ \hline
sign $v_{p}$ & $-$ & $0$ & $+$ & $+$ \\ \hline
horizon & $\left\vert v_{p}\right\vert <1$ & $0$ & $\left\vert
v_{p}\right\vert <1$ & $\left\vert v_{p}\right\vert =1$ \\ \hline
\end{tabular}

For the Schwarzschild metric, $\left\vert v_{p}\right\vert $ monotonically
decreases from infinity to singularity.

Table 2. The behavior of $\left\vert v_{p}\right\vert $ from infinity to the
inner horizon for the RN metric.

\begin{tabular}{|l|l|}
\hline
& $\left\vert v_{p}\right\vert $ \\ \hline
From infinity to $r_{0}$ & decreases \\ \hline
From $r_{0}$ to $r_{-}$ & increases \\ \hline
\end{tabular}

\section{Kinematics of collisions in terms of peculiar velocities}

If two particles collide, one can define the energy in their centre of mass
frame in the point of collision as $E_{c.m.}^{2}=-P_{\mu }P^{\mu }$, where $%
P^{\mu }$ is the total four-momentum in this point. Then,%
\begin{equation}
E_{c.m.}^{2}=m_{1}^{2}+m_{2}^{2}+2m_{1}m_{2}\gamma \text{,}  \label{ecm}
\end{equation}%
where $\gamma =-u_{1\mu }u^{2\mu }$ is the effective Lorentz factor of
relative motion. Outside the horizon, 
\begin{equation}
\gamma =\frac{\varepsilon _{1}\varepsilon _{2}-\sigma Z_{1}Z_{2}}{f}\text{,}
\label{2}
\end{equation}%
$\sigma =\sigma _{1}\sigma _{2}$.

The problem is that outside the horizon, both particles have $\varepsilon
_{1,2}>0$, so for $\sigma =+1$ the BSW effect is absent, $E_{c.m.}$ remains
modest. This can be seen clearly from (\ref{2}) if one expands the numerator
in the Taylor series with respect to small $f$ and retains the main
contribution.

For $\sigma =-1$, the BSW effect is possible formally but it cannot be
realized physically near a black hole since in this case it is difficult to
create a particle moving near the horizon in the opposite direction. (See
also Sec. IV A of \cite{fraq} about details.)\ Instead, one can consider
collision near white Schwarzchild-like holes \cite{gpwhite}. However, we
concentrate on a more realistic case of a black hole and collisions in the
contracting $T-$region \cite{nov}. In this region, the counterpart of eq. (%
\ref{2}) obtained from (\ref{muu}) reads%
\begin{equation}
\gamma =\frac{Z_{1}Z_{2}-\varepsilon _{1}\varepsilon _{2}}{g}\text{.}
\label{gat}
\end{equation}

Here, the sign is absorbed by the factors $\varepsilon _{1,2}$, so we do not
write $\sigma $ explicitly.

The above formulas can be rewritten in terms of velocities. In particular,

\begin{equation}
\gamma \equiv \frac{1}{\sqrt{1-w^{2}}}=\gamma _{1}\gamma _{2}-\sigma \sqrt{%
\gamma _{1}^{2}-1}\sqrt{\gamma _{2}^{2}-1}\text{,}  \label{1}
\end{equation}%
where $\gamma _{a}=\frac{1}{\sqrt{1-V_{a}^{2}}}$ ($a=1,2$) are individual
gamma factors, $w$ has the meaning of the relative velocity of two
particles. In derivation of (\ref{1}) from (\ref{gat}) we used (\ref{ein}).
Then%
\begin{equation}
\gamma =\frac{1-\sigma V_{1}V_{2}}{\sqrt{1-V_{1}^{2}}\sqrt{1-V_{2}^{2}}},
\label{VV}
\end{equation}

where now $V$ is given by eq. (\ref{vez}).

Using (\ref{vV}), one obtains $\gamma $ in terms of peculiar velocities:

\begin{equation}
\gamma =\frac{1-v_{p1}v_{p2}}{\sqrt{1-v_{p1}^{2}}\sqrt{1-v_{p2}^{2}}}.
\label{gap}
\end{equation}

This formula looks exactly like in special relativity (SR). This shows the
reason to use the "river model of black hole" -- we can exploit standard SR
formulae for the local processes (decay, scattering, etc.) if we insert
kinematical values taken with respect to GP frame, which have been found
from general relativity (GR) formulae.

\section{High energy collisions of massive particles}

In what follows, we will use terminology accepted in the literature on the
BSW effect \cite{prd}. Applying it in our context, we call the particle with 
$\varepsilon =0$ critical and particles with $\varepsilon \neq 0$ usual.

\subsection{Collision between critical and usual particles}

As was shown earlier, the BSW effect occurs when the critical particle
collides with a usual one at the horizon. For critical particle 1, $%
V_{1}<1\,\ $remains separated from $1$ near the horizon, for usual particle
2 $V_{2}\rightarrow 1$ . In other words, from the viewpoint of a stationary
observer, a rapid particle hits a slow target. But in terms of peculiar
velocities the situation turns out to be opposite. Indeed, according to (\ref%
{vpe}), for the critical particle $v_{p}\rightarrow 1$ and for a usual
particle $v_{p}\neq 1$. Thus the GP system gives description of the BSW
effect complimentary to the stationary frame. It is also interesting that on
the horizon $v_{p}$ can have either sign, so in this frame both particles
can move in any direction with respect to each other.

To be more concrete, let us consider the scenario in which particle 1 is
critical, so $\varepsilon _{1}=0$. Then, according to (\ref{Zz}), $Z_{1}=%
\sqrt{g}$. We assume that a usual particle 2 has $\varepsilon _{2}>1$ and
comes from infinity (from the right $R-$region \cite{nov}). Taking into
account (\ref{vv}), (\ref{1-e}) and (\ref{gap}), we obtain for collisions
under the horizon in the $T$ region: 
\begin{equation}
\gamma =\frac{\sqrt{1+g}+\left\vert v_{p2}\right\vert }{\sqrt{1-v_{p2}^{2}}%
\sqrt{g}}.  \label{gaf}
\end{equation}%
Near the horizon, $g\rightarrow 0$, $v\approx 1$. Then, it follows from (\ref%
{gaf}), (\ref{eps}) that%
\begin{equation}
\gamma \approx \frac{\left\vert \varepsilon _{2}\right\vert }{\sqrt{g}}\gg 1%
\text{,}  \label{gae2}
\end{equation}

It is worth stressing that although $v_{p}\approx 1$ for the critical
particle when $g\rightarrow 0$, it cannot be equal to $1$ exactly since this
would imply that $g=0$. Then, a timelike particle would became lightlike one
that is impossible. Correspondingly, $\gamma $ is unbounded but cannot be
infinite. This is one more manifestation of the principle of cosmic
censorship \cite{cens}.

\subsection{Head-on collision between usual particles}

There is also another version of high energy collisions near the horizon
when both particles are usual but move in opposite directions ($\sigma =-1$)
thus experiencing head-on collision \cite{pir1}, \cite{pir3}. This means
that radial momenta have the opposite signs. It is worth stressing that
under the horizon such a momentum is related not to $\dot{r}$ which is
negative for all particles but to $\dot{y}$ according to (\ref{my}).
According to (\ref{2}) with $\sigma =-1$, head-on collision of any two
particles gives rise to the unbounded growth of $\gamma $ and $E_{c.m.}$, no
fine-tuning of particle parameters is required. In particular, in our metric
this can be realized with collisions between particles with $\varepsilon >0$
and $\varepsilon <0$. In the "river model" we get the same conclusion from
the SR formula (\ref{gap}), if we remember that $v_{p}$\ for $\varepsilon >0$%
\ tends to some subluminal value (\ref{vpe}) while $v_{p}$\ for the particle
with $\varepsilon <0$\ tends to unity (\ref{vhor}).

We saw that high energy conditions under discussion required participation
of particles with $\varepsilon \leq 0$. But such particles do not exist in
the outer region. How can they appear under the event horizon? One option
consists in that the corresponding trajectories are given "by hand" as
geodesics that come from the left $R$ region. However, in a more physically
realistic situation, we can assume a black hole to be smoothly accreted by
surrounding matter. Then, the initial conditions are given at infinity (from
which a particle falls into a black hole) where $\varepsilon >0$. It means
that a particle should experience an additional collision in the inner
region to achieve the required state or create a new particle with $%
\varepsilon \leq 0$. For a pure radial motion, according to Eqs. (\ref{vv})
and (\ref{sign}) this requires%
\begin{equation}
v_{p}\geq \frac{1}{v}\text{.}
\end{equation}%
Afterwards, this new particle collides with one more particle coming from
infinity. It is worth noting that the present discussion concerning
properties of particles inside the event horizon, does not depend on whether
the Schwarzschild metric inside is a part of the eternal black-white hole or
a result of collapse.

The whole scenario depends on the type of a black hole and the horizon. Let
an initial particle decay inside the Schwarzschild black hole. Then, a new
particle with $\varepsilon \leq 0$ cannot return to the horizon, so any new
collisions will have finite $E_{c.m.}$ Let us try another option in which
decay occurs immediately after crossing the horizon. Is it possible to
create the critical particle in this process? The problem is that if
critical particle 1 and usual particle 2 meet in the same point near the
horizon, their $E_{c.m.}$ is divergent according to general rues \cite{prd}.
However, if particle 0 has a finite mass, this is impossible (as is seen
easily in the frame comoving to particle 0).

We conclude that for the Schwarzschild black hole it is impossible to
achieve unbounded $E_{c.m.}$ if initial conditions include only particle
coming form infinity. But this is possible if the critical trajectory
existed inside a black hole from the very beginning.

The situation near the inner horizon of the RN metric is more interesting
and rich but its detailed description is beyond the scope of the present
article and will be done elsewhere.

\section{Collision of a massive and massless particles\label{mm}}

Let us consider radial motion of a photon and observer, so $L=0=l$. Let a
free falling observer emit or receive a photon. Its frequency measured by
this observer is equal to 
\begin{equation}
\omega =-k_{\mu }u^{\mu }.  \label{om}
\end{equation}

Outside the horizon, talking into account (\ref{mt}) - (\ref{mr}), (\ref{kf}%
), (\ref{kl}) we find:%
\begin{equation}
\frac{\omega }{\omega _{0}}=\frac{\varepsilon -\sigma Z}{f}\text{.}
\label{wzq}
\end{equation}

Here, $\sigma =+1$ if both objects (an observer and a photon) move in the
same direction and $\sigma =-1$ if they do this in the opposite ones.

This expression can be simplified further in terms of peculiar velocities.
It follows from (\ref{eps}), (\ref{zvp}) that%
\begin{equation}
\frac{\omega }{\omega _{0}}=\frac{1}{1+\sigma v}\sqrt{\frac{1+\sigma v_{p}}{%
1-\sigma v_{p}}}\text{.}  \label{out}
\end{equation}

Under the horizon, the analogue of formula (\ref{wzq}) gives us 
\begin{equation}
\frac{\omega }{\omega _{0}}=\frac{Z-\sigma \left\vert \varepsilon
\right\vert }{g}\text{,}  \label{rad}
\end{equation}%
$\omega _{0}>0$, $\sigma =\pm 1$.

If $\sigma =-1$, $\omega \rightarrow \infty $ when $g\rightarrow 0$. Let now 
$\sigma =+1$. It is seen from (\ref{zf}), (\ref{zf2}) that if $\varepsilon
\neq 0$ (usual particle), $Z-\left\vert \varepsilon \right\vert =O(g)$,
hence $\frac{\omega }{\omega _{0}}$ remains bounded near the horizon.
However, if $\varepsilon =0$ (the critical particle), $Z=\sqrt{g\text{ }}$
according to (\ref{Zz}),%
\begin{equation}
\frac{\omega }{\omega _{0}}=\frac{1}{\sqrt{g}}.  \label{red}
\end{equation}

When $r\rightarrow r_{+}$, $g\rightarrow 0$ and we again obtain the
counterpart of the BSW effect (infinite blueshift in a given case).

In terms of peculiar velocities, we obtain similarily to (\ref{out}), 
\begin{equation}
\frac{\omega }{\omega _{0}}=\frac{1}{v+\alpha }\sqrt{\frac{1+\alpha v_{p}}{%
1-\alpha v_{p}}}\text{,}  \label{al}
\end{equation}%
where now $\alpha =sign\sigma \varepsilon $.

\section{Redshift vs. blueshift in the river model}

The results for redshift (blueshift) caused by collisions of massive and
massless particles (\ref{out}) and (\ref{al}) contain the peculiar
velocities and thus are obtained in the "river model". In this Seciton, we
discuss their properties in more detail. We use the standard definition of
the redshift $1+z=\omega _{0}/\omega $, so infinite blueshift ($\omega
=\infty $) corresponds to $1+z=0$. If the emitter is at rest with respect to
the GP frame, and the observer moves with the velocity $v_{p}$, the
resulting redshift can be decomposed into two parts

\begin{equation}
1+z=(1+z_{g})(1+z_{D})  \label{zz}
\end{equation}%
the same way as the cosmological redshifts in the case of non-zero velocity
with respect to Hubble flow (see, for example, \cite{Davis}). Here $z_{g}$
is a gravitational redshift which would be detected by the same observer if
it has zero peculiar velocity, and $z_{D}$ is the standard relativistic
Doppler shift caused by non-zero $v_{p}$ of the \textit{observer,} 
\begin{equation}
1+z_{D}=\sqrt{(1-v_{p})/(1+v_{p})}.  \label{zd}
\end{equation}%
Gravitational redshift has been found in our previous paper \cite{3} where
it is shown that if we denote by $v_{e}$ and $v_{o}$ correspondingly the
values of free fall velocity $v$ of emitter and observer, than for the
inward photon 
\begin{equation}
1+z_{g}=(1+v_{o})/(1+v_{e})  \label{zg}
\end{equation}%
(see eq.(28) of that paper). This means that for an infinitely remote source
at rest ($v_{e}=0$) the "classical" formula $z_{g}=v_{o}$ is still valid in
a black hole background. Thus $z_{g}$ is always positive (a redshift), and
is finite everywhere except for the singularity. In particular, $z_{g}=1$ at
any horizon.

On the other hand, the Doppler shift observed by $\varepsilon \leq 0$
particle diverges since $v_{p}\sim 1$. Note also that since the peculiar
velocity in this case is positive and, thus, directed outward, the redshift
is negative (so, it is actually a blueshift). The resulting frequency change
can be found by combining these two effects according to (\ref{zz}). For the
emitter at rest ($v_{e}=0$) it coincides with (\ref{out}) as it should be.
For the critical particle $\varepsilon =0$ we have from (\ref{vv}) and (\ref%
{zz}) - (\ref{zg}) with $v_{e}=0$ that%
\begin{equation}
1+z=v_{p}^{-1}\sqrt{1-v_{p}^{2}}\text{.}  \label{zcr}
\end{equation}

When a particle in question approaches the horizon, $v_{p}\rightarrow 1$ and 
$1+z\rightarrow 0$, so we have an infinite blueshift. It appears as a result
of combination of a finite gravitational redshift and the diverging Doppler
blueshift.

Let us now consider an observer with $\varepsilon <0$. For a particular case
of the RN black hole this situation is described in monographs and even
textbooks (see, e.g. \cite{chandra}, page 214, \cite{he}, page 161). The
well-known pecularity of this case is that the observer in question will be
able to see all future of our Universe, and that is why this case is often
described in popular science. Near the inner horizon, such a particle
experiences head-on collision with a photon coming from infinity, and, as
the $v_{p}$ of the observer tends to unity, we have the resulting blueshift 
\begin{equation}
1+z=(1+v_o)\sqrt{(1-v_{p})/(1+v_{p})}  \label{1+z}
\end{equation}%
diverging in the same way as for the $\varepsilon =0$ case. This is in sharp
contrast with a situation when a particle crossing the horizon emits or
absorbs a photon of a finite local frequency just on the horizon. Then, the
resulting blueshift is finite -- see details in \cite{along}.

Note also, that any object free falling from the outer $R$ region has $%
\varepsilon >0$. Let it cross the inner horizon. Then, in (\ref{1+z}) $%
v_{o}=1$ according to (\ref{vf}). Using also Eq. (\ref{vpe}) we see that the
redshift for such an observer when he/she crosses the inner horizon is
finite and is equal to $z=2\varepsilon -1$. Meanwhile, the whole picture of
an observer moving inside a black hole is more complex. Here, we would only
like to indicate, preliminarily, some important features. This observer is
being beaten by more and more energetic photons while travelling closer to
the inner horizon. Moreover, he would experience more and more energetic
collisions with free-falling accreting matter consisting of massive
particles present in any realistic situation. The energy of such collisions
diverges at the inner horizon. The role of such collisions is not
investigated in literature and was only briefly mentioned in the end of Ref. 
\cite{inner}. We postpone more detailed consideration for a separate work.

\section{Summary, conclusions and perspectives}

We considered the geodesic trajectories with $\varepsilon =0$ existing under
the event horizon. In addition to known earlier zero angular momentum
observers \cite{72} and zero energy ones \cite{01}, \cite{02}, the present
class of zero radial momentum observers gives one more example of the fact
that strong gravity opens a window of interesting possibilities and new
kinds of trajectories absent in the flat space-time.

We traced how the concept of the river of space works, introducing peculiar
velocities and reformulating kinematics of particles in their terms
including the BSW effect or its analogues. From the methodological point of
view this is interesting because the procedure we used is well known in
cosmology, where the motion of a distant object is commonly considered as a
sum of Hubble flow and peculiar velocities. When local processes in a
distant galaxy or cluster are considered, only peculiar velocities are
relevant, moreover (as far as we are interested in local events, but not the
quantities detected by a distant observer) we can apply standard special
relativity formulae to objects moving with given peculiar velocities.
General relativity is needed in this approach only to find evolution of
peculiar velocities in time, after that SR can be applied.

As it was already shown in \cite{ham}, a similar concept can be developed
for black hole space-times, if we work in the Gullstrand-Painlev\'{e} frame.
It appears possible in this frame to decompose a general geodesic motion of
a particle falling into a black hole into a sum of the flow and a peculiar
velocity, such that local kinematics of physical processes can be found
using peculiar velocities by standard SR formulae. In our present paper we
systematically studied the properties of peculiar velocities and applied
them to describe analogs of BSW processes and other known effects, like
infinite blueshift seen by an observer crossing the inner horizon. The use
of peculiar velocities gives us a description of these effects complimentary
to the standard one. In some cases it is easier it understand these effects
in the approach under discussion that gives a more intuitively clear
picture. The price for this is less known properties of peculiar velocities
in comparison with velocities with respect to stationary coordinates. For
example, in the outer region of black hole gravity obviously accelerates a
free-fall of particles in our situation (where all particles outside are
usual), though it slows down the motion with respect to Gullstrand-Painlev%
\'{e} frame! However, advantages to use SR in describing physical events
deep inside a horizon is a reasonable motivation to study the possibilities
given by "river model of black hole" in a systematic way.

In the present work, we mainly restricted ourselves by the Schwarzchild
metric and focused on the properties of trajectories with $\varepsilon =0$
or small $\varepsilon $. Meanwhile, such trajectories exist for any
nonextremal black hole including the RN one, so the results are quite
generic in this sense. We only cursorily touched upon the case of the RN
metric in what concerns general kinematic properties of a particle in terms
of peculiar velocity.

The full dynamics of matter and field behavior inside the RN metric is much
more complicated. It includes the influence of wave propagation on the inner
horizon \cite{grs1}, \cite{grs2}, creation of particle by the electric field 
\cite{sn}, mass "inflation" \cite{mi}, etc. It would be interesting to
consider how massive particles discussed in our work may collide with very
large energies before back reaction of created particles on the space-time
metric becomes important, and how the corresponding processes affect the
properties of the Cauchy horizon. However, this is a separate issue that
requires further investigation for which we consider our work as a first
step.

\section*{Acknowledgements}

The work was supported by the Russian Government Program of Competitive
Growth of Kazan Federal University. O. Z. also thanks for support SFFR,
Ukraine, Project No. 32367.

\end{document}